\documentclass[prd,preprint,superscriptaddress,amsmath,amssymb,nofootinbib]{revtex4}
\usepackage{graphicx}
\usepackage{dcolumn}
\usepackage{bm}
\usepackage{amssymb}
\usepackage{amsmath}
\usepackage{epsfig}    
\usepackage{color}
\usepackage{slashed}
\usepackage{hhline}

\def\be{\begin{equation}}
\def\ee{\end{equation}}
\newcommand{\bea}{\begin{eqnarray}}
\newcommand{\eea}{\end{eqnarray}}
\newcommand{\nn}{\nonumber}

\begin{document}

 \begin{flushright} {KIAS-P20039, APCTP Pre2020 - 013}  \end{flushright}

\title{Modular $A_4$ symmetric inverse seesaw model\\
 with $SU(2)_L$ multiplet fields }

\author{Takaaki Nomura}
\email{nomura@kias.re.kr}
\affiliation{School of Physics, KIAS, Seoul 02455, Republic of Korea}

\author{Hiroshi Okada}
\email{hiroshi.okada@apctp.org}
\affiliation{Asia Pacific Center for Theoretical Physics (APCTP) - Headquarters San 31, Hyoja-dong,
Nam-gu, Pohang 790-784, Korea}
\affiliation{Department of Physics, Pohang University of Science and Technology, Pohang 37673, Republic of Korea}

\date{\today}

\begin{abstract}
 We propose an inverse seesaw model with large $SU(2)_L$ multiplets applying modular $A_4$ symmetry where 
 $SU(2)_L$ quartet and septet fermions are introduced as triplets under the symmetry.
 The neutral components of the quartet contribute to mass matrix for inverse seesaw mechanism and interactions involving the septet induce Majorana mass terms of these extra fermions. 
 Although there are several free parameters in the mass matrix we can obtain some predictions for observables in neutrino sector such as CP phases, sum of neutrino mass and effective mass for neutrinoless double beta decay, especially around fixed points of modulus motivated by string theories.
 \end{abstract}
\maketitle

\section{Introduction}

A mechanism to generate neutrino mass is one of the most open questions which require physics beyond the standard model (SM).
In constructing a model of neutrino mass generation, flavor symmetry would play an important role to control flavor structure.
A model with flavor symmetry would give predictions for observables in neutrino sector.

One of interesting possibilities is application of modular flavor symmetries proposed by~\cite{Feruglio:2017spp, deAdelhartToorop:2011re} in constructing a neutrino mass model. 
In the framework, a coupling can be transformed under a non-trivial representation of a non-Abelian discrete group and we can restrict flavor structure of neutrino mass matrix.  
Then some typical groups are found to be available in this framework such as $A_4$~\cite{Feruglio:2017spp, Criado:2018thu, Kobayashi:2018scp, Okada:2018yrn, Nomura:2019jxj, Okada:2019uoy, deAnda:2018ecu, Novichkov:2018yse, Nomura:2019yft, Okada:2019mjf,Ding:2019zxk,Nomura:2019lnr,Kobayashi:2019xvz,Asaka:2019vev,Zhang:2019ngf, Gui-JunDing:2019wap,Kobayashi:2019gtp,Nomura:2019xsb,Behera:2020sfe,Wang:2019xbo,Okada:2020dmb,Okada:2020rjb,Nomura:2020opk}, $S_3$ \cite{Kobayashi:2018vbk, Kobayashi:2018wkl, Kobayashi:2019rzp, Okada:2019xqk}, $S_4$ \cite{Penedo:2018nmg, Novichkov:2018ovf, Kobayashi:2019mna,King:2019vhv,Okada:2019lzv,Criado:2019tzk,Wang:2019ovr,Wang:2020dbp}, $A_5$ \cite{Novichkov:2018nkm, Ding:2019xna,Criado:2019tzk}, larger groups~\cite{Baur:2019kwi}, multiple modular symmetries~\cite{deMedeirosVarzielas:2019cyj}, and double covering of $A_4$~\cite{Liu:2019khw} and $S_4$~\cite{Novichkov:2020eep, Liu:2020akv} in which observables like masses, mixings, and CP phases for quarks and/or leptons are predicted.~\footnote{Some reviews are useful for understanding the non-Abelian group and its applications to flavor structure~\cite{Altarelli:2010gt, Ishimori:2010au, Ishimori:2012zz, Hernandez:2012ra, King:2013eh, King:2014nza, King:2017guk, Petcov:2017ggy}.}
Moreover, CP violation in models with modular symmetry is discussed in ref.~\cite{Kobayashi:2019uyt,Novichkov:2019sqv},
a systematic approach to understand the origin of CP transformations has been discussed in ref.~\cite{Baur:2019iai},  
a possible correction from K\"ahler potential is also discussed in ref.~\cite{Chen:2019ewa}, 
and cases of half integral modular weight is discussed in ref.~\cite{Liu:2020msy}.

We apply a modular $A_4$ symmetry to inverse seesaw model with large $SU(2)_L$ multiplets introduced in ref.~\cite{Nomura:2018cfu}.
In our approach, $SU(2)_L$ quartet and septet fermions are introduced as triplets under the modular $A_4$ symmetry
where the neutral components of the quartet contribute to mass matrix for inverse seesaw mechanism~\cite{Mohapatra:1986bd, Wyler:1982dd} and interactions involving the septet induce Majorana mass matrix. 
Although there are several free parameters in the mass matrix, we can obtain some predictions for observables in neutrino sector such as CP phases, sum of neutrino mass and effective mass for neutrinoless double beta decay.
Especially, we find regions around fixed points of modulus motivated by string theories.

This letter is organized as follows.
In Sec. II, {we review our model and formulate the lepton sector. 
Then we discuss phenomenologies of neutrinos. In Sec. III we discuss an extra charged particles at collider experiments.
 Finally we devote the summary of our results and the conclusion.}

\section{Model setup and Constraints}
\begin{table}[t!]
\begin{tabular}{|c||c|c|c|c||c|c|c|}\hline\hline  
& ~$L_L^a$~& ~$e_R^a$~& ~$\psi^a$~& ~$\Sigma_R^a$~& ~$H_2$~& ~$H_4$~& ~$H_5$~\\\hline\hline 
$SU(2)_L$   & $\bm{2}$  & $\bm{1}$  & $\bm{4}$  & $\bm{7}$ & $\bm{2}$   & $\bm{4}$ & $\bm{5}$   \\\hline 
$U(1)_Y$    & $-\frac12$  & $-1$ & {-$\frac32$}  & $0$  & $\frac12$ & {$\frac32$}  &{$2$} \\\hline
$A_4$  & $\bm{1,1',1''}$  & $\bm{1,1',1''}$  & $\bm{3}$  & $\bm{3}$ & $\bm{1}$   & $\bm{1}$ & $\bm{1}$   \\\hline 
$-k$    & $0$  & $0$ & $-2$  & $0$  & $0$ & {$0$}  &{$0$} \\\hline
\end{tabular}
\caption{Charge assignments of the  lepton and scalar fields
under $SU(2)_L\times U(1)_Y$, where the upper index $a$ is the number of family that runs over 1-3 and
all of them are singlet under $SU(3)_C$. }\label{tab:1}
\end{table}

\begin{center} 
\begin{table}[t]
\begin{tabular}{|c||c|c|c|c|c|}\hline
{Yukawa coupling}  & ~{ $A_4$}~& ~$-k_I$~     \\\hline 
{ $\bm{\tilde Y_{1,1',1''}^{(4)}}$} & ${\bf 1,1',1''}$ & ${\bf 4}$      \\\hline
{ $\bm{Y^{(2)}_3}$} & ${\bf 3}$ & ${\bf 2}$      \\\hline
{ $\bm{\tilde Y^{(4)}_3}$} & ${\bf 3}$ & ${\bf 4}$      \\\hline
\end{tabular}
\caption{Modular weight assignment for Yukawa coupling $\bm{Y}$ and its transformation under $A_4$ for giving flavor structure of different neutral fermion mass matrices.}
\label{tab:couplings}
\end{table}
\end{center}

In this section we show our model with modular $A_4$ symmetry.
For the fermion sector, we introduce three families of vector-like fermions $\psi$ with $(\bm{4},-3/2)$ charge under the $SU(2)_L\times U(1)_Y$ gauge symmetries, and right-handed fermions $\Sigma_R$ with $(\bm{7},0)$ charge under the same gauge symmetry, all of which are given as triplets of $A_4$.
%
For the scalar sector, we add quartet and quintet scalar fields $H_4$ and $H_5$ which have respectively $3/2$ and $2$ charges under the $U(1)_Y$ gauge symmetry, where SM-like Higgs field is identified as $H_2$.
These scalar fields develop vacuum expectation values{(VEVs)} denoted by $\langle H_i\rangle\equiv v_{i}/\sqrt2$ ($i=2,4,5$) inducing spontaneous electroweak symmetry breaking.
All the field contents and their assignments under the symmetry are summarized in Table~\ref{tab:1}, where the quark sector is exactly the same as the one of the SM and omitted.

The  modular forms of weight 2, {$Y^{(2)}_{\bm3} = (y_{1},y_{2},y_{3})$},  transforming as a triplet of $A_4$ are given by Dedekind eta-function  $\eta(\tau)$ and its derivative $\eta'(\tau)$~\cite{Feruglio:2017spp}:
\begin{eqnarray} 
\label{eq:Y-A4}
y_{1}(\tau) &=& \frac{i}{2\pi}\left( \frac{\eta'(\tau/3)}{\eta(\tau/3)}  +\frac{\eta'((\tau +1)/3)}{\eta((\tau+1)/3)}  
+\frac{\eta'((\tau +2)/3)}{\eta((\tau+2)/3)} - \frac{27\eta'(3\tau)}{\eta(3\tau)}  \right), \nonumber \\
y_{2}(\tau) &=& \frac{-i}{\pi}\left( \frac{\eta'(\tau/3)}{\eta(\tau/3)}  +\omega^2\frac{\eta'((\tau +1)/3)}{\eta((\tau+1)/3)}  
+\omega \frac{\eta'((\tau +2)/3)}{\eta((\tau+2)/3)}  \right) , \label{eq:Yi} \\ 
y_{3}(\tau) &=& \frac{-i}{\pi}\left( \frac{\eta'(\tau/3)}{\eta(\tau/3)}  +\omega\frac{\eta'((\tau +1)/3)}{\eta((\tau+1)/3)}  
+\omega^2 \frac{\eta'((\tau +2)/3)}{\eta((\tau+2)/3)}  \right)\,.
\nonumber
\end{eqnarray}
Modular forms with higher weight are constructed by the products of $Y^{(2)}_{\bm 3}$.
Here we also summarize modular forms to construct mass and Yukawa interaction terms in Table~\ref{tab:couplings}.
Note that $\tilde Y_{(4)}$'s are modular forms constructed by $Y_{\bm 3}^{(2)}$ and its conjugate such that 
\begin{align}
&\tilde Y^{(4)}_{\bm 1} = [Y^{(2)}_{\bm 3} Y^{(2)*}_{\bm 3}]_{\bm 1} =y_1 y_1^* + y_2 y_2^* + y_3y_3^* \equiv \tilde y_{S},\\
&\tilde Y^{(4)}_{\bm 1'} = [Y^{(2)}_{\bm 3} Y^{(2)*}_{\bm 3}]_{\bm 1'} =y_3 y_2^* + y_1 y_3^* + y_2y_1^* \equiv \tilde y_{S'},\\
&\tilde Y^{(4)}_{\bm 1''} = [Y^{(2)}_{\bm 3} Y^{(2)*}_{\bm 3}]_{\bm 1''} =y_2 y_3^* + y_1 y_2^* + y_3y_1^* \equiv \tilde y_{S''},\\
& \tilde Y^{(4)}_{\bf3} = [Y^{(2)}_{\bm 3} Y^{(2)*}_{\bm 3}]_{\bm 3 (\rm sym)} =
\left[\begin{array}{c}
2 y_1 y_1^*-y_2y_2^* -y_3y_3^* \\ 
2 y_3 y_2^* -y_1y_3^* -y_2y_1^* \\ 
2 y_2 y_3^* -y_1 y_2^* -y_3 y_1^* \\ 
\end{array}\right]
\equiv \left[\begin{array}{c}
\tilde y_{T_1} \\ 
\tilde y_{T_2} \\ 
\tilde y_{T_3} \\ 
\end{array}\right] \\
& \tilde Y'^{(4)}_{\bf3} = [Y^{(2)}_{\bm 3} Y^{(2)*}_{\bm 3}]_{\bm 3 (\rm anti-sym)} =
\left[\begin{array}{c}
 y_2 y_2^*-y_3y_3^* \\ 
 y_1 y_3^* -y_2y_1^* \\ 
y_3 y_1^* -y_1 y_2^* \\ 
\end{array}\right]
\equiv \left[\begin{array}{c}
\tilde y'_{T_1} \\ 
\tilde y'_{T_2} \\ 
\tilde y'_{T_3} \\ 
\end{array}\right],
\end{align}
where subscripts "sym" and "anti-sym" indicate symmetric and anti-symmetric product of two ${\bf 3}$ representation.
Under modular $A_4$ symmetry, interaction term is invariant when it is $A_4$ trivial singlet and sum of modular weights is zero. 
Then we write the renormalizable Yukawa Lagrangian under these symmetries as follows
\begin{align}
-{\cal L_\ell}
& =  y_{\ell_{aa}} \bar L^a_L H_2 e^a_R  +  {\bm Y_{3}^{(2)*}} [ \bar L_L  \psi^c_L H_5^*]  \nn\\
& +  {\bm Y_{3}^{(2)*}} [\bar \psi_L \Sigma_R  H_4^*]_{\bm 3}
+  {\bm Y_{3}^{(2)}}[(\bar\psi^c_R)  \Sigma_R H_4]_{\bm 3} \nn\\
&+M_1 {\bm \tilde Y_{3}^{(4)}} [\bar \psi_R \psi_L]_{\bm3} +M_2 {\bm \tilde Y'^{(4)}_{3}} [\bar \psi_R \psi_L]_{\bm3} +M_3 {\bm \tilde Y_{\{\bm1\}}^{(4)}} [\bar \psi_R \psi_L]_{\{\bm1\}}+ M_{\Sigma_{} }[\bar \Sigma^{c}_R \Sigma_R]_{\bm1}
+ {\rm h.c.}, \label{eq:lag}
\end{align}
where ${\{\bm 1\}}\equiv(\bm{1,1',1''})$,  $SU(2)_L$ index is omitted assuming it is contracted to be gauge invariant inside square bracket, subscript for square bracket indicates $A_4$ representation for corresponding operator, and upper indices ($a,b=$1--3) are the number of families. 
Note here that $y_\ell$ is diagonal thanks to the feature of $A_4$ symmetry.
We can explicitly write flavor structure of these terms expanding $A_4$ representations as follows.
For the second term of Eq.~\eqref{eq:lag} we obtain
\begin{align}
{\bm Y_{3}^{(2)*}} [ \bar L_L  \psi^c_L H_5^*] = & \left[ a_D \bar L_L^1 ({\bm Y_{3}^{(2)*}} \psi^c_L)_{{\bm 1}} +b_D \bar L_L^2 ({\bm Y_{3}^{(2)*}} \psi^c_L )_{{\bm 1'}} +c_D \bar L_L^3 ({\bm Y_{3}^{(2)*}} \psi^c_L)_{{\bm 1''}}  \right] H_5^* \nn \\
 = & \left[ a_D \bar L_L^1 (y_1^*  (\psi^{c }_L)^1  + y_3^*  (\psi^{c }_L)^2 + y_2^*  (\psi^{c }_L)^3  ) + b_D \bar L_L^2 (y_2^*  (\psi^{c }_L)^2  + y_1^*  (\psi^{c }_L)^3 + y_3^*  (\psi^{c }_L)^1  )  \right. \nn \\
 & \left. + c_D \bar L_L^3 (y_3^*  (\psi^{c }_L)^3  + y_1^*  (\psi^{c }_L)^2 + y_2^*  (\psi^{c }_L)^1  ) \right] H_5^*,
 \label{eq:mass1}
\end{align}
where $\{a_D,b_D,c_D \}$ are free parameters and invariance under $SU(2)_L$ is implicitly imposed as we assume the invariance for the other terms.
For the fourth term of Eq.~\eqref{eq:lag} we obtain 
\begin{align}
&{\bm Y_{3}^{(2)}}[(\bar\psi^c_R)  \Sigma_R H_4]_{\bm 3} \nn \\
& =   {\bm Y_{3}^{(2)}}[ d(\bar\psi^c_R  \Sigma_R)_{\rm sym} + e(\bar\psi^c_R  \Sigma_R)_{\rm anti-sym} ] H_4 \nn \\
& =  \left[ \frac{d}{3} \{ y_1(2  (\bar\psi^c_R)^1  \Sigma_R^1 - (\bar\psi^c_R)^2 \Sigma_R^3 - (\bar\psi^c_R)^3  \Sigma_R^2 ) + y_2(2  (\bar\psi^c_R)^2  \Sigma_R^2 - (\bar\psi^c_R)^1 \Sigma_R^3 - (\bar\psi^c_R)^3  \Sigma_R^1 )   \right. \nn \\
& \qquad + y_3(2  (\bar\psi^c_R)^3  \Sigma_R^3 - (\bar\psi^c_R)^1 \Sigma_R^2 - (\bar\psi^c_R)^2  \Sigma_R^1 ) \} \nn \\
& \left. \qquad + \frac{e}{2} \{ y_1 ((\bar\psi^c_R)^2 \Sigma_R^3 - (\bar\psi^c_R)^3  \Sigma_R^2) + y_2 ((\bar\psi^c_R)^3 \Sigma_R^1 - (\bar\psi^c_R)^1  \Sigma_R^3) + y_3 ((\bar\psi^c_R)^1 \Sigma_R^2 - (\bar\psi^c_R)^2  \Sigma_R^1) \} \right] H_4.
 \label{eq:mass2}
\end{align}
The fifth term of Eq.~\eqref{eq:lag} gives 
\begin{align}
& M_1 {\bm \tilde Y_{3}^{(4)}} [\bar \psi_R \psi_L]_{\bm3} \nn \\
& = M_1 {\bm \tilde Y_{3}^{(4)}} [ c_1 (\bar \psi_R \psi_L)_{\rm sym} +  c_2 (\bar \psi_R \psi_L)_{\rm anti-sym}  ] \nn \\
& = M_0 \left[ \rho \{ \tilde y_{T_1} (2  (\bar\psi_R)^1  \psi_L^1 - (\bar\psi_R)^3 \psi_L^3 - (\bar\psi_R)^2 \psi_L^2 ) 
+ \tilde y_{T_2}(2  (\bar\psi_R)^3 \psi_L^2 - (\bar\psi_R)^1 \psi_L^3 - (\bar\psi_R)^2 \psi_L^1 ) \right. \nn \\
& \qquad + \tilde y_{T_3} (2  (\bar\psi_R)^2 \psi_L^3 - (\bar\psi_R)^1 \psi_L^2 - (\bar\psi_R)^3 \psi_L^1 ) \} \nn \\
& \left. \qquad + \gamma' \{ \tilde y_{T_1} ((\bar\psi_R)^3 \psi_L^3 - (\bar\psi_R)^2  \psi_L^2) + \tilde y_{T_2} ((\bar\psi_R)^2 \psi_L^1 - (\bar\psi_R)^1  \psi_L^3) + \tilde y_{T_3} ((\bar\psi_R)^1 \psi_L^2 - (\bar\psi^c)^3  \psi_R^1) \} \right]
 \label{eq:mass3}
\end{align}
where we redefined parameters at third line.
Similarly the sixth term of Eq.~\eqref{eq:lag} gives 
\begin{align}
& M_2 {\bm \tilde Y'^{(4)}_{3}} [\bar \psi_R \psi_L]_{\bm3} \nn \\
& = M_2 {\bm \tilde Y'^{(4)}_{3}} [ c'_1 (\bar \psi_R \psi_L)_{\rm sym} +  c'_2 (\bar \psi_R \psi_L)_{\rm anti-sym}  ] \nn \\
& = M_0 \left[ \delta \{ \tilde y'_{T_1} (2  (\bar\psi_R)^1  \psi_L^1 - (\bar\psi_R)^3 \psi_L^3 - (\bar\psi_R)^2 \psi_L^2 ) + \tilde y'_{T_2}(2  (\bar\psi_R)^3  \psi_L^2 - (\bar\psi_R)^1 \psi_L^3 - (\bar\psi_R)^2  \psi_L^1 )   \right. \nn \\
& \qquad + \tilde y'_{T_3} (2  (\bar\psi_R)^2  \psi_L^3 - (\bar\psi_R)^1 \psi_L^2 - (\bar\psi_R)^3  \psi_L^1 ) \} \nn \\
& \left. \qquad + \sigma \{ \tilde y'_{T_1} ((\bar\psi_R)^3 \psi_L^3 - (\bar\psi_R)^2  \psi_L^2) + \tilde y'_{T_2} ((\bar\psi_R)^2 \psi_L^1 - (\bar\psi_R)^1  \psi_L^3) + \tilde y'_{T_3} ((\bar\psi_R)^1 \psi_L^2 - (\bar\psi^c)^3  \psi_R^1) \} \right].
 \label{eq:mass4}
\end{align}
The seventh term of Eq.~\eqref{eq:lag} then provides
\begin{align}
& M_3 {\bm \tilde Y_{\{\bm1\}}^{(4)}} [\bar \psi_R \psi_L]_{\{\bm1\}} \nn \\
& = M_0 \left[  {\bm \tilde Y_{\bm 1}^{(4)}} [\bar \psi_R \psi_L]_{\bm 1} + \beta {\bm \tilde Y_{\bm 1'}^{(4)}} [\bar \psi_R \psi_L]_{\bm 1''} + \gamma {\bm \tilde Y_{\bm 1''}^{(4)}} [\bar \psi_R \psi_L]_{\bm 1'}  \right] \nn \\
& = M_0 \left[ \tilde y_{S_{1}} (\bar \psi_R^1 \psi_L^1 + \bar \psi_R^3 \psi_L^3 + \bar \psi_R^2 \psi_L^2) + \beta \tilde y_{S_{1'}} (\bar \psi_R^3 \psi_L^2 + \bar \psi_R^1 \psi_L^3 + \bar \psi_R^2 \psi_L^1) + \right. \nn \\
& \left. \qquad \qquad + \tilde y_{S_{1''}} (\bar \psi_R^2 \psi_L^3 + \bar \psi_R^1 \psi_L^2 + \bar \psi_R^3 \psi_L^1) \right],
 \label{eq:mass5}
\end{align}
where we redefined parameter in second line.
After spontaneous symmetry breaking we obtain mass matrices $m_\ell=y_\ell v/\sqrt2$, $m_D \propto {\bm Y_{3_{ab}}^{(2)*}} v_5$, $m_R\propto{\bm Y_{3_{ab}}^{(2)}} v_4$, and $m_L\propto{\bm Y_{3_{ab}}^{(2)*}} v_4$.   

{
\noindent \underline{\it Scalar potential and VEVs}:
We write the scalar potential in our model as follows
\begin{align}
{\cal V} = & -\mu_h^2 |H_2|^2 + M_4^2 |H_4|^2 + M_5^2 |H_5|^2  + \lambda_H |H_2|^4 \nonumber \\
& + \mu [H_4 H_2  H_5^* +h.c] + \lambda_0 [H_4^* H_2 H_2 H_2 + h.c.] \nonumber \\
& + {\cal V}_{\rm trivial},
\end{align}
where ${\cal V}_{\rm trivial}$ indicates other trivial 4-point terms and $SU(2)_L$ indices are implicitly contracted in the second line to be gauge invariant.
Applying vacuum condition $\partial {\cal V}/\partial v_i = 0$, we obtain the VEVs such that
\begin{align}
v_2 \sim \sqrt{\frac{\mu^2_h}{\lambda_H}}, \quad v_4 \sim \frac{\lambda_0 v^3 }{M_4^2}, \quad v_5 \sim \frac{\mu v_4 v}{M_5^2},
\end{align}
where we have used VEV hierarchy of $v_4, v_5 \ll v_2$.
Thus $v_4$ and $v_5$ can be naturally $\mathcal{O}(1)$ GeV scale if $M_4$ and $M_5$ are around TeV scale.
}

\noindent \underline{\it $\rho$ parameter}:
The VEVs of $H_4$ and $H_5$ are restricted since it shift $\rho$-parameter from $1$ at tree level: 
\begin{align}
\rho\approx \frac{v_2^2+7 v_4^2+10 v_5^2}{v_2^2+  v_4^2 + 2 v_5^2},
\end{align}
where the experimental value is given by $\rho=1.0004^{+0.0003}_{-0.0004}$ at $2\sigma$ confidence level~\cite{pdg}.
On the other hand, it is required that $v_{SM}=\sqrt{v_2^2+7 v_4^2+10 v_5^2}\simeq v_2\approx$246 GeV. 
Therefore $v_4$ and $v_5$ are restricted to be small to satisfy the constraint of $\rho$ parameter.
Hereafter, we assume these VEVs to be $v_2\approx$245.9 GeV, $v_4\approx$1.67 GeV, and $v_5\approx$1.72 GeV, which are typical scale for the VEVs satisfying the constraint.
%

\noindent \underline{\it Exotic particles} :
The scalars and fermions with large $SU(2)_L$ multiplet contain exotic charged particles 
where we can write multiplets in terms of components as
\begin{align}
& H_4 = (\phi_4^{+++}, \phi_4^{++}, \phi_4^{+}, \phi_4^0)^T, \\
& H_5 = (\phi_5^{++++}, \phi_5^{+++}, \phi_5^{++}, \phi_5^{+}, \phi_5^0)^T, \\
& \psi_{L(R)} = (\psi^{0}, \psi^{-}, \psi^{--}, \psi^{---})^T_{L(R)}, \\
& \Sigma_R = (\Sigma^{+++}, \Sigma^{++}, \Sigma^{+}, \Sigma^0, \Sigma^-, \Sigma^{--}, \Sigma^{---})_R^T. 
\end{align}
The masses of components in scalar multiplets $H_4$ and $H_5$ are respectively given by $\sim M_4$ and $\sim M_5$ since $v_{4,5} \ll M_{4,5}$.
The charged components in the quartet $\psi_{L(R)}$ have Dirac mass $M$ and mass terms for neutral component are discussed with neutrino sector below.
The mass of septet fermion is given by $M_\Sigma$ where charged components have Dirac mass term constructed by pairs of positive-negative charged components inside the multiplet.
Notice that charged particles in the same multiplet have degenerate mass at tree level which will be shifted at loop level by order of few GeV~\cite{Cirelli:2005uq}.

\begin{figure}[tb]
\begin{center}
 \includegraphics[width=10.0cm]{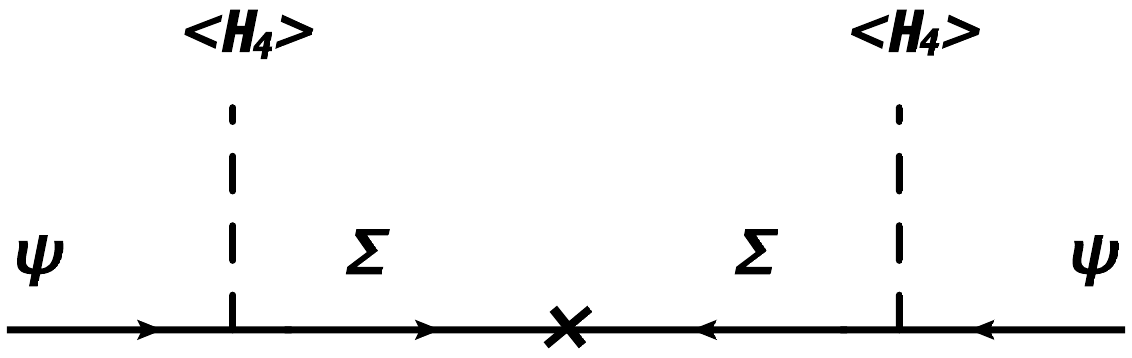}
\caption{Feynman diagram to generate the Majorana masses of $\mu_{L/R}$.}
\label{fig:mu_mass}
\end{center}\end{figure}

\noindent \underline{\it Neutrino sector}:
After the spontaneously symmetry breaking, we obtain neutral fermion mass matrix in basis of $(\nu_L,\psi_R^{0c},\psi_L^0)^T$ as follows
\begin{align}
M_N
&=
\left[\begin{array}{ccc}
0 & 0 & m_D^*  \\ 
0 & \mu_R^* & M \\ 
m_D^\dag  & M^T & \mu_L^* \\ 
\end{array}\right],
\end{align}
where $M$ consists of $M_{1,2}$ in Eq.(\ref{eq:lag}), $\mu_{R}$ is given by $ m_{R}{\cal M}_\Sigma^{-1}  m^T_{R}$ on the analogical manner of seesaw mechanism~\footnote{Since $\mu_L$ does not contribute to the neutrino mass, we do not consider it.},
as shown in Fig.~\ref{fig:mu_mass}.
Each mass matrix is explicitly derived from Eqs.~\eqref{eq:mass1}--\eqref{eq:mass5} such that
\begin{align}
m_D&\equiv \frac{\sqrt3 v_5}{2} \tilde m_D=\frac{\sqrt3 v_5}{2}
\left[\begin{array}{ccc}
a_D & 0 & 0  \\ 
0 & b_D & 0 \\ 
0  & 0 & c_D \\ 
\end{array}\right]
\left[\begin{array}{ccc}
y_1^* & y_2^* & y_3^*  \\ 
y_3^* & y_1^* & y_2^*  \\ 
y_2^* & y_3^* & y_1^*  \\ 
\end{array}\right]
\left[\begin{array}{ccc}
1 & 0 & 0  \\ 
0 & 0 & 1  \\ 
0 & 1 & 0  \\ 
\end{array}\right]
 ,\\
{\cal M}_\Sigma&\equiv M_\Sigma\tilde {\cal M}_\Sigma= M_\Sigma
\left[\begin{array}{ccc}
1 & 0 & 0  \\ 
0 & 0 & 1 \\ 
0  & 1 & 0 \\ 
\end{array}\right],\\
m_R&\equiv \frac{\sqrt3 v_4}{2\sqrt5}\tilde m_R=\frac{\sqrt3 v_4}{2\sqrt5}
\left[\begin{array}{ccc}
\frac{2d}{3}y_1 & -\frac{d}{3}y_3+\frac{e}{2}y_3 & -\frac{d}{3}y_2-\frac{e}{2}y_2  \\ 
-\frac{d}{3}y_3-\frac{e}{2}y_3 & \frac{2d}{3}y_2 & -\frac{d}{3}y_1+\frac{e}{2}y_1 \\ 
-\frac{d}{3}y_2+\frac{e}{2}y_2  & -\frac{d}{3}y_1- \frac{e}{2}y_1 & \frac{2d}{3}y_3 \\ 
\end{array}\right] ,\
 \\
\label{eq:M}
M&\equiv M_0 \tilde M=M_0
\left[\begin{array}{ccc}
1 & 0 & 0  \\ 
0 & 0 & 1  \\ 
0 & 1 & 0  \\ 
\end{array}\right]
\left(
\left[\begin{array}{ccc}
\tilde y_{S_1} & \gamma\tilde y_{S_{1''}} & \beta\tilde y_{S_{1'}}  \\ 
 \gamma\tilde y_{S_{1''}} & \beta\tilde y_{S_{1'}} & \tilde y_{S_{1}}  \\ 
 \beta\tilde y_{S_{1'}} & \tilde y_{S_{1}} & \gamma\tilde y_{S_{1''}}  \\ 
\end{array}\right]\right.\\
&+\left.
\left[\begin{array}{ccc}
2(\rho \tilde y_{T_1}+\delta\tilde y'_{T_1}) & (-\rho+\gamma')\tilde y_{T_3}+(\sigma-\delta) \tilde y'_{T_3}
&  -(\rho+\gamma')\tilde y_{T_2}-(\sigma+\delta) \tilde y'_{T_2}  \\ 
-(\rho+\gamma')\tilde y_{T_3}-(\sigma+\delta) \tilde y'_{T_3}  & 2(\rho \tilde y_{T_2}+\delta\tilde y'_{T_2}) & 
(-\rho+\gamma')\tilde y_{T_1}+(\sigma-\delta) \tilde y'_{T_1}  \\ 
(-\rho+\gamma')\tilde y_{T_2}+(\sigma-\delta) \tilde y'_{T_2}&  -(\rho+\gamma')\tilde y_{T_1}-(\sigma+\delta) \tilde y'_{T_1}  & 2(\rho \tilde y_{T_3}+\delta\tilde y'_{T_3}) \\ 
\end{array}\right]\right),\nn 
\end{align}
where $\tilde m_D,\tilde {\cal M}_\Sigma, \tilde m_R,\tilde M$ are defined by dimensionless parameters.
Since $v_{4,5} \ll v$ and $\{M, M_\Sigma\}$ is TeV scale we can naturally realize hierarchy of mass matrix element $\mu_R \ll m_D \ll M$ for inverse seesaw mechanism~\footnote{Such hierarchies could be also explained by several mechanisms such as radiative models~\cite{Dev:2012sg, Dev:2012bd, Das:2017ski} and effective models with higher order terms \cite{Okada:2012np}.}. 
Then the active neutrino mass matrix can approximately be found as follows
\begin{align}
m_\nu&\approx m_D^* M_{}^{-1} ( m_{R}{\cal M}_\Sigma^{-1}  m^T_{R})^* (M_{}^T)^{-1} m_D^\dag\nn\\
&=\frac{27}{80}\frac{v_4^2v_5^2}{M_0^2 M_\Sigma}\tilde m_D^*\tilde M^{-1} \tilde m_R^*\tilde{\cal M}_\Sigma^{-1} \tilde m_R^\dag (\tilde M^T)^{-1} \tilde m_D^\dag, \nn \\
& \equiv \kappa \tilde m_\nu,
\label{eq:Mnu}
\end{align}
where $\tilde m_\nu$ is given by dimensionless matrices.
Then, $\tilde m_\nu$ is diagonalized by applying a unitary matrix as $V^\dag_\nu (\tilde m_\nu^\dag \tilde m_\nu)V_\nu=(\tilde D_{\nu_1}^2,\tilde D_{\nu_2}^2,\tilde D_{\nu_3}^2)$. 
In this case, mass dimension parameter $\kappa$ is determined by 
\begin{align}
(NO):\  \kappa^2= \frac{|\Delta m_{\rm atm}^2|}{\tilde D_{\nu_3}^2-\tilde D_{\nu_1}^2},
\quad
(IO):\  \kappa^2= \frac{|\Delta m_{\rm atm}^2|}{\tilde D_{\nu_2}^2-\tilde D_{\nu_3}^2},
 \end{align}
where  $\Delta m_{\rm atm}^2$ is atmospheric neutrino mass difference squares and $\kappa\equiv \frac{27}{80}\frac{v_4^2v_5^2}{M_0^2 M_\Sigma}$ from Eq.~\eqref{eq:Mnu}. 
Subsequently, the solar mass different squares can be written in terms of $\kappa$ such as
\begin{align}
\Delta m_{\rm sol}^2= {\kappa^2}({\tilde D_{\nu_2}^2-\tilde D_{\nu_1}^2}),
 \end{align}
 where we obtain the value as output in our numerical analysis and it should be compared with the observed value.
For heavy sterile neutrino, we obtain pseudo Dirac mass for $\mu_0 \ll M$ and mass eigenvalues are obtained by diagonalizing $M$ where 
we write these eigenvalues as $M_{1,2,3}$ which will be output in our numerically analysis.
  
Here one finds $U_{PMNS}=V_\nu$ since the charged-lepton is originally in diagonal basis. Then, 
it is parametrized by three mixing angle $\theta_{ij} (i,j=1,2,3; i < j)$, one CP violating Dirac phase $\delta_{CP}$,
and two Majorana phases $\{\alpha_{21}, \alpha_{32}\}$:
\begin{equation}
U_{PMNS} = 
\begin{pmatrix} c_{12} c_{13} & s_{12} c_{13} & s_{13} e^{-i \delta_{CP}} \\ 
-s_{12} c_{23} - c_{12} s_{23} s_{13} e^{i \delta_{CP}} & c_{12} c_{23} - s_{12} s_{23} s_{13} e^{i \delta_{CP}} & s_{23} c_{13} \\
s_{12} s_{23} - c_{12} c_{23} s_{13} e^{i \delta_{CP}} & -c_{12} s_{23} - s_{12} c_{23} s_{13} e^{i \delta_{CP}} & c_{23} c_{13} 
\end{pmatrix}
\begin{pmatrix} 1 & 0 & 0 \\ 0 & e^{i \frac{\alpha_{21}}{2}} & 0 \\ 0 & 0 & e^{i \frac{\alpha_{31}}{2}} \end{pmatrix},
\end{equation}
where $c_{ij}$ and $s_{ij}$ stands for $\cos \theta_{ij}$ and $\sin \theta_{ij}$ respectively. 
These mixings are given in terms of the component of $U_{PMNS}$ as follows:
\begin{align}
\sin^2\theta_{13}=|(U_{PMNS})_{13}|^2,\quad 
\sin^2\theta_{23}=\frac{|(U_{PMNS})_{23}|^2}{1-|(U_{PMNS})_{13}|^2},\quad 
\sin^2\theta_{12}=\frac{|(U_{PMNS})_{12}|^2}{1-|(U_{PMNS})_{13}|^2}.
\end{align}
In addition we can compute the Jarlskog invariant, $\delta_{CP}$ from PMNS matrix elements $U_{\alpha i}$:
\begin{equation}
J_{CP} = \text{Im} [U_{e1} U_{\mu 2} U_{e 2}^* U_{\mu 1}^*] = s_{23} c_{23} s_{12} c_{12} s_{13} c^2_{13} \sin \delta_{CP},
\end{equation}
and the Majorana phases are also estimated in terms of other invariants $I_1$ and $I_2$ constructed by PMNS matrix elements:
\begin{equation}
I_1 = \text{Im}[U^*_{e1} U_{e2}] = c_{12} s_{12} c_{13}^2 \sin \left( \frac{\alpha_{21}}{2} \right), \
I_2 = \text{Im}[U^*_{e1} U_{e3}] = c_{12} s_{13} c_{13} \sin \left( \frac{\alpha_{31}}{2}  - \delta_{CP} \right).
\end{equation}
Furthermore, the effective mass for the neutrinoless double beta decay is given by
\begin{align}
\langle m_{ee}\rangle=\kappa|\tilde D_{\nu_1} c^2_{12} c^2_{13}+\tilde D_{\nu_2} s^2_{12} c^2_{13}e^{i\alpha_{21}}+\tilde D_{\nu_3} s^2_{13}e^{i(\alpha_{31}-2\delta_{CP})}|,
\end{align}
where its observed value could be measured by KamLAND-Zen experiment in future~\cite{KamLAND-Zen:2016pfg}. 
In our numerical analysis, we will adopt the neutrino experimental data at 3$\sigma$ interval~\cite{Esteban:2018azc, Nufit} as follows:
\begin{align}
&{\rm NO}: \Delta m^2_{\rm atm}=[2.431, 2.622]\times 10^{-3}\ {\rm eV}^2,\
\Delta m^2_{\rm sol}=[6.79, 8.01]\times 10^{-5}\ {\rm eV}^2,\\
&\sin^2\theta_{13}=[0.02044, 0.02437],\ 
\sin^2\theta_{23}=[0.428, 0.624],\ 
\sin^2\theta_{12}=[0.275, 0.350],\nn\\
&{\rm IO}: \Delta m^2_{\rm atm}=[2.413, 2.606]\times 10^{-3}\ {\rm eV}^2,\
\Delta m^2_{\rm sol}=[6.79, 8.01]\times 10^{-5}\ {\rm eV}^2,\\
&\sin^2\theta_{13}=[0.02067, 0.02461],\ 
\sin^2\theta_{23}=[0.433, 0.623],\ 
\sin^2\theta_{12}=[0.275, 0.350],\nn
\end{align}
where NO and IO stand for normal and inverted ordering respectively.

\subsection{Non-unitarity}
Here, let us briefly discuss non-unitarity matrix $U'_{MNS}$ due to the existence of sterile neutrinos.
This is typically parametrized by the following form: 
\begin{align}
U'_{MNS}\equiv \left(1-\frac12 FF^\dag\right) U_{MNS},
\end{align}
where $F\equiv  (M^{T}_{})^{-1} m_D$ is a hermitian matrix, and $U'_{MNS}$ represents the deviation from the unitarity. 
The global constraints are found via several experimental results such as the SM $W$ boson mass $M_W$, the effective Weinberg angle $\theta_W$, several ratios of $Z$ boson fermionic decays, invisible decay of $Z$, electroweak universality, measured Cabbibo-Kobayashi-Maskawa, and lepton flavor violations~\cite{Fernandez-Martinez:2016lgt};
the resulting constraint is then given by~\cite{Agostinho:2017wfs}
\begin{align}
|FF^\dag|\le  
\left[\begin{array}{ccc} 
2.5\times 10^{-3} & 2.4\times 10^{-5}  & 2.7\times 10^{-3}  \\
2.4\times 10^{-5}  & 4.0\times 10^{-4}  & 1.2\times 10^{-3}  \\
2.7\times 10^{-3}  & 1.2\times 10^{-3}  & 5.6\times 10^{-3} \\
 \end{array}\right].
\end{align} 
In our model, $F\equiv  (M^{T}_{})^{-1} m_D= \frac{\sqrt3}{2} \frac{v_5}{M_0}(\tilde M^T)^{-1}\tilde m_D$ if $\tilde M$ and $\tilde m_D$ are taken to be the same order.
Therefore, Non-unitarity can be controlled by $ \frac{v_5}{M_0}$ which is naturally small due to the constraint from the rho parameter.

\section{Numerical analysis and phenomenology}

In this section, we carry out numerical analysis of neutrino sector searching for allowed parameters fitting neutrino data.
Then we discuss phenomenology in the model.

\subsection{Numerical analysis of neutrino sector}

In our numerical analysis we scan free parameters such that 
\begin{equation}
\{ a_D, b_D, c_D, e, d, \beta, \gamma, \rho, \delta, \sigma, \gamma' \} \in [0.1, 1], \quad {\rm Re}[\tau] \in [0,0.5], \quad {\rm Im}[\tau] \in [0.5, 2],\label{eq:ip-rg}
\end{equation}
where combination of massive parameters in Eq.~(\ref{eq:Mnu}) is determined from neutrino mass scale.

\begin{figure}[tb]\begin{center}
\includegraphics[width=80mm]{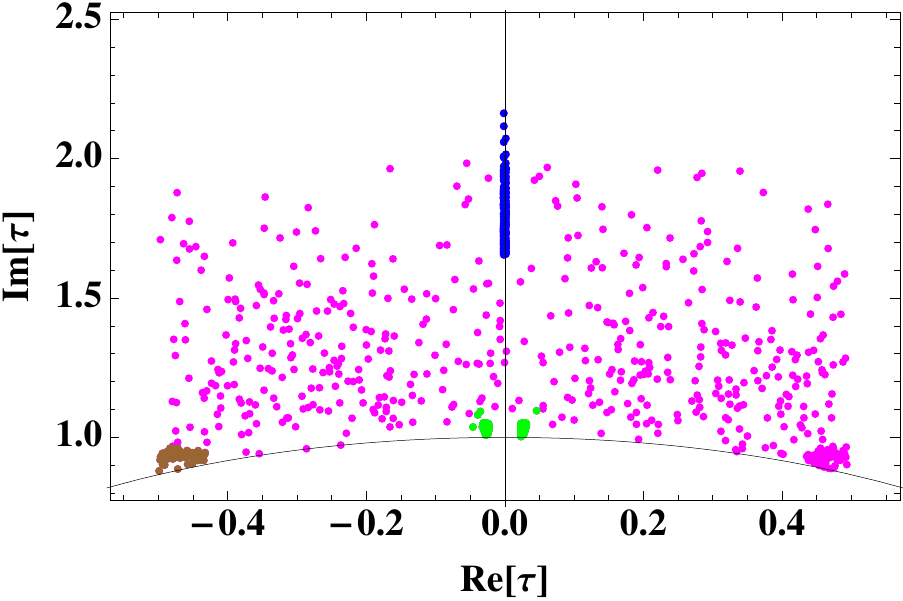}
\caption{The allowed region for $\tau$ in case of NO,  where the magenta shows whole the range, the blue points correspond to the fixed point $\tau=i\times\infty$, the green points correspond to the fixed point $\tau=i$, and the brown points correspond to the fixed point $\tau=w$.} 
\label{fig:tau}
\end{center}\end{figure}

In Fig.~\ref{fig:tau}, we demonstrate the correlation plot between Re$[\tau]$ and Im$[\tau]$ in the fundamental region in case of NO, where the magenta shows whole the range, the blue points correspond to the fixed point $\tau=i\times\infty$, the green points correspond to the fixed point $\tau=i$, and the brown points correspond to the fixed point $\tau=\exp(2\pi i/3) (\equiv w)$. Both of fixed points are especially favored by string theory~\cite{Kobayashi:2020uaj}, since these provide the minimum potential.
{\it Here, we focus on searching the allowed region in case of NO only, since we have found that there are not any region at around the fixed points after the global analysis. Therefore, IO would not have any predictions in the lepton sector. }

\begin{figure}[tb]\begin{center}
\includegraphics[width=80mm]{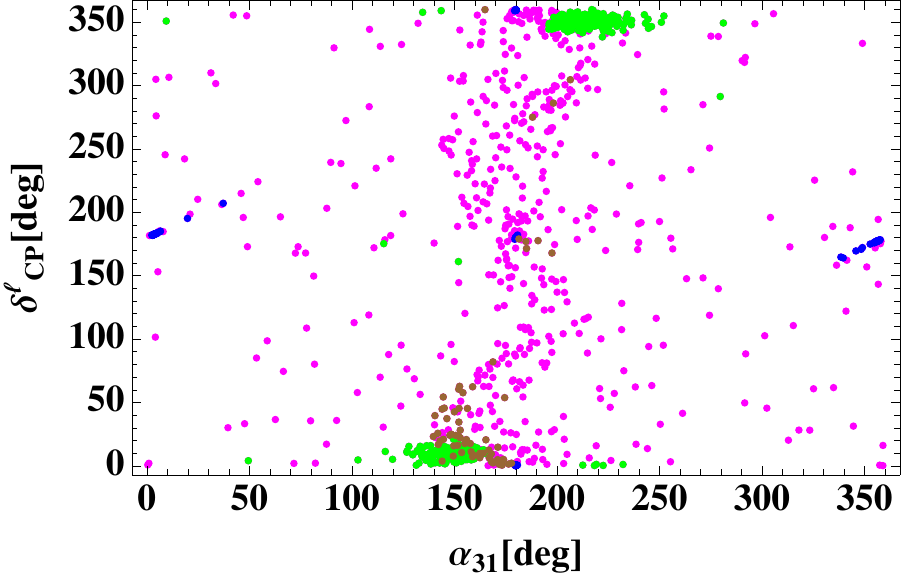} 
\caption{The correlation plot of Dirac CP Phase $\delta^\ell_{CP}$ and $\alpha_{31}$ in case of NO,
where the color represents the same as Fig.\ref{fig:tau}.} 
\label{fig:majodir-NH}
\end{center}\end{figure}

In Fig.~\ref{fig:majodir-NH}, we demonstrate the correlation plot of Dirac CP Phase $\delta^\ell_{CP}$ and $\alpha_{31}$ in case of NO.
Although the magenta region would have a dense tendency, whole the region would be allowed.
 Once we focus on the fixed points, we can predict these values as follows.
Among the region of fixed point $\tau=i \times \infty$ that is colored by blue,  $\alpha_{31}=[(-20^\circ) - 50^\circ],180^\circ$, and $\delta^\ell_{CP}=0^\circ,[160^\circ- 210^\circ]$.
Among the region of fixed point $\tau=i$ that is colored by green,  $\alpha_{31}=[100^\circ-250^\circ]$, and $\delta^\ell_{CP}=0^\circ,[(-10^\circ) - 20^\circ]$.
Among the region of fixed point $\tau=w$ that is colored by brown,  $\alpha_{31}=[150^\circ - 200^\circ]$, and $\delta^\ell_{CP}= [0^\circ-50^\circ]$.

\begin{figure}[tb]\begin{center}
\includegraphics[width=80mm]{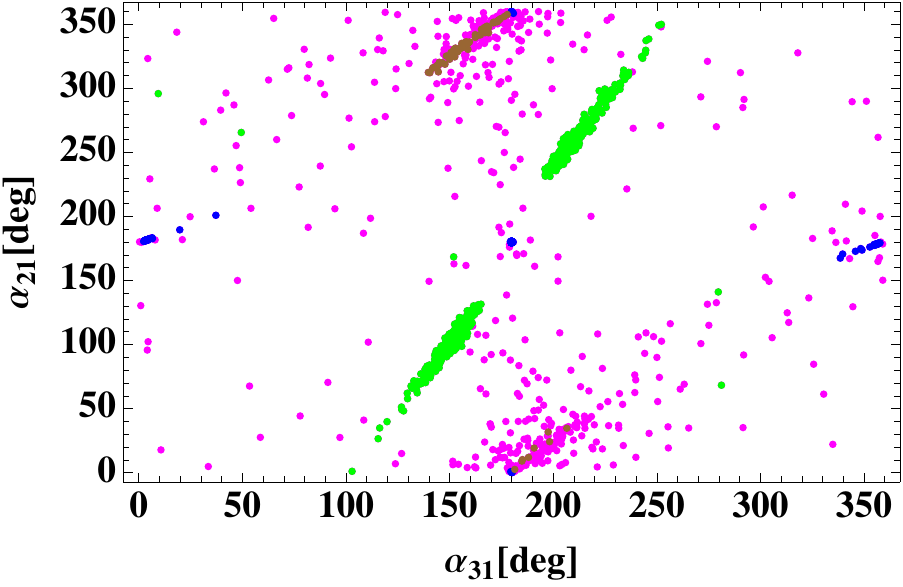} \
\caption{The correlation plot of $\alpha_{21}$ and $\alpha_{31}$ in case of NO,
where the color represents the same as Fig.\ref{fig:tau}.}  
\label{fig:majos-NH}\end{center}\end{figure}

In Fig.~\ref{fig:majos-NH}, we show the correlation plot between $\alpha_{21}$ and $\alpha_{31}$ in case of NO.
Although the magenta region would have a dense tendency, whole the region would be allowed.
 Once we focus on the fixed points, we can predict something below.
Among the region of fixed point $\tau=i \times \infty$, $\alpha_{21},[170^\circ$-$200^\circ]$.
Among the region of fixed point $\tau=i$, $\alpha_{21}=[50^\circ-130^\circ]$.
Among the region of fixed point $\tau=w$, $\alpha_{21}=[(-50^\circ)-40^\circ]$.

\begin{figure}[h!]\begin{center}
\includegraphics[width=80mm]{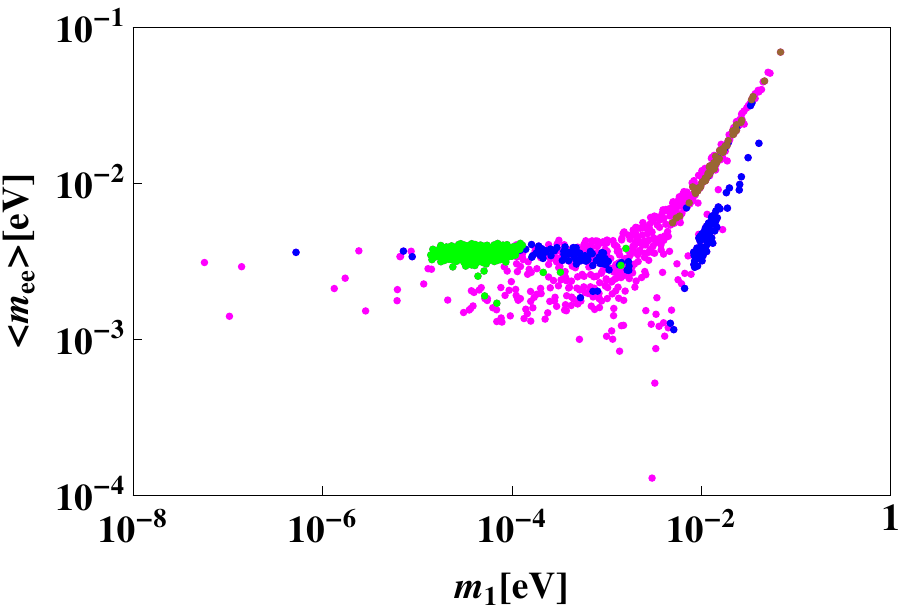} \
\caption{The lightest neutrino mass $m_1$ versus the effective mass for the neutrinoless double beta decay $\langle m_{ee}\rangle$ in case of NO,
where the color represents the same as Fig.\ref{fig:tau}.}
\label{fig:m1mee-NH}
\end{center}\end{figure}

%
In Fig.~\ref{fig:m1mee-NH}, we show the effective mass for the neutrinoless double beta decay $\langle m_{ee}\rangle$ as a function of the lightest neutrino mass $m_1$in case of NO. 
It suggests that $\langle m_{ee}\rangle$ and $m_1$ respectively cover the range of $[10^{-4}-0.1]$ eV and  $[10^{-8}-0.1]$ eV in the left figure, where each the lowest bound would come from our choice of the minimum input parameters $0.1$. 
If we focus on the fixed point $\tau=i$, these allowed regions are specified by $m_1\sim{\cal O}(10^{-4})$ eV, and $\langle m_{ee}\rangle\sim{\cal O}(3\times10^{-3})$ eV.
For fixed point $\tau = w$, we obtain $\langle m_{ee}\rangle \simeq m_1 \gtrsim 10^{-2} $ eV.

\begin{figure}[h!]\begin{center}
\includegraphics[width=80mm]{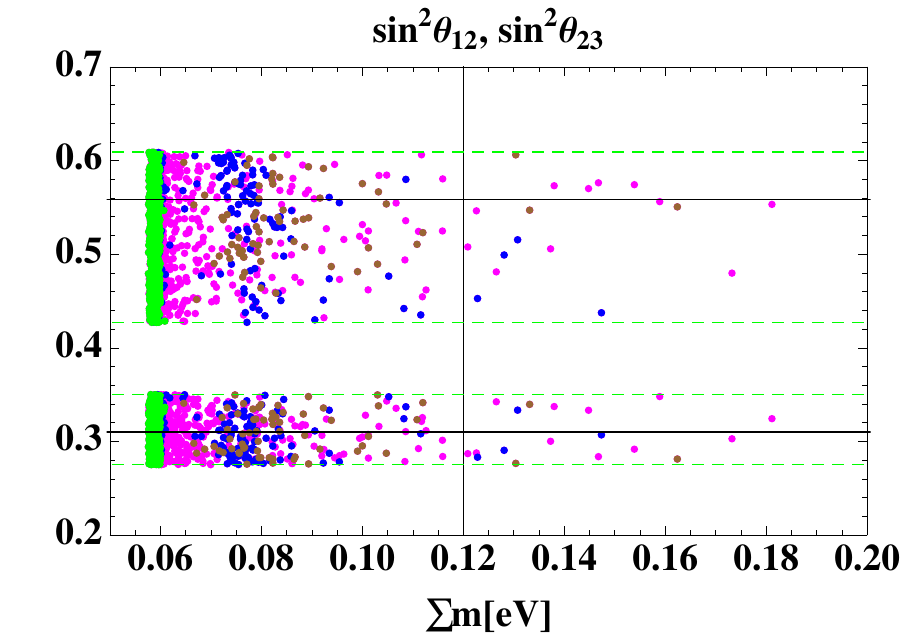}\
\includegraphics[width=80mm]{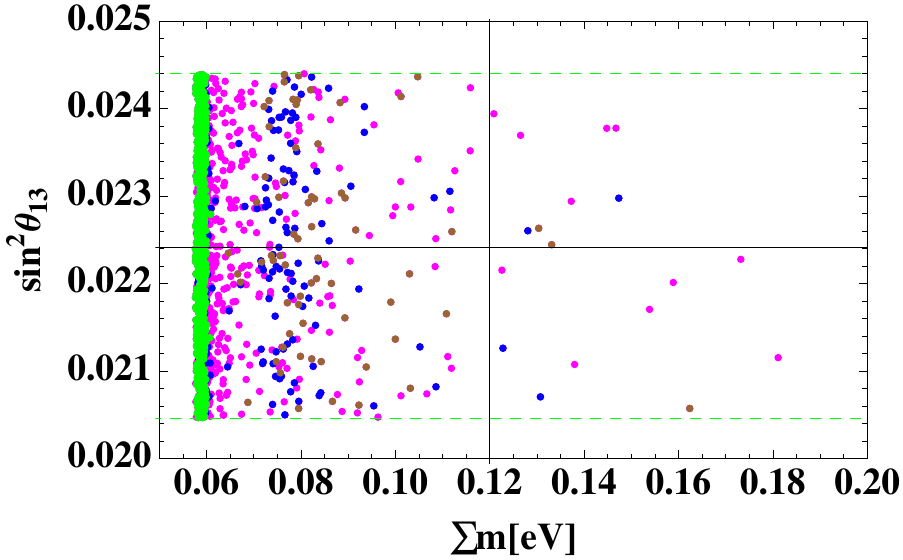}\
\caption{The sum of neutrino masses $\sum m$ versus $\sin^2\theta_{12}(red), \sin^2\theta_{23}(blue)$ for the left figure and $\sin^2\theta_{13}$ for right figure in case of NO,
where the color represents the same as Fig.\ref{fig:tau}.}   
\label{sum1223-NH}\end{center}\end{figure}

%
In Fig.~\ref{sum1223-NH}, we show three mixing angles of PMNS as functions of sum of neutrino masses $\sum m$.
All the mixings $s_{12},s_{23},s_{13}$ cover all the experimental ranges within 3$\sigma$ interval,
while we find $\sum m=[0.058-0.18]$ eV which is almost within the region allowed by the cosmological constraint.
The minimum value of $\sum m$ also comes from  our choice of the minimum input parameters 0.1.  
The fixed point $\tau=i$ favors the region $\sum \sim0.06$ eV.

\subsection{Collider phenomenology}
In this subsection, we briefly discuss collider phenomenology of the model focusing on signals from 
production of charged fermion $\psi^-$ in quartet $\psi$ since it gives a unique signal related to neutrino mass generation mechanism.
At the Large Hadron Collider(LHC) $\psi^\pm$ can be produced via gauge interaction $pp \to \gamma/Z \to \psi^- \psi^+$.
The singly charged fermion $\psi^-$ decays into $\ell^+ \phi^{--}_5$ through Yukawa interaction related to neutrino mass generation:
\begin{equation}
{\bm Y_{3}^{(2)*}} [ \bar L_L  \psi^c_L H_5^*] + h.c. \supset {\bm Y_{3}^{(2)*}} \left[ \frac{1}{\sqrt{2}} \bar \ell_L  (\psi^-)^c_L \phi_5^{--} - \frac{\sqrt{3}}{2} \bar \nu_L  (\psi^-)^c_L \phi_5^{-} \right].
\end{equation}
Thus flavor dependence of branching ratio(BR) depends on modular form and it is restricted by after fitting neutrino data.
Assuming mass of $\Phi_5$ is lighter than that of $\psi$, doubly charge scalar $\phi_5^{\pm \pm}$ dominantly decays into $W^\pm W^\pm$ through gauge interaction
\begin{equation}
(D_\mu \Phi_5)^\dagger (D^\mu \Phi_5) \supset \sqrt{3} v_5 g_2^2 W^\pm_\mu W^{\pm \mu} \phi^{\mp \mp}_5,
\end{equation}
where $g_2$ is gauge coupling of $SU(2)_L$.
Then we obtain signal of leptons with W bosons as $pp \to \psi^- \psi^+ \to \phi^{++}_5 \phi^{--}_5 \ell^+ \ell'^- \to W^+W^+ W^- W^- \ell^+ \ell'^-$.

\begin{figure}[t]\begin{center}
\includegraphics[width=80mm]{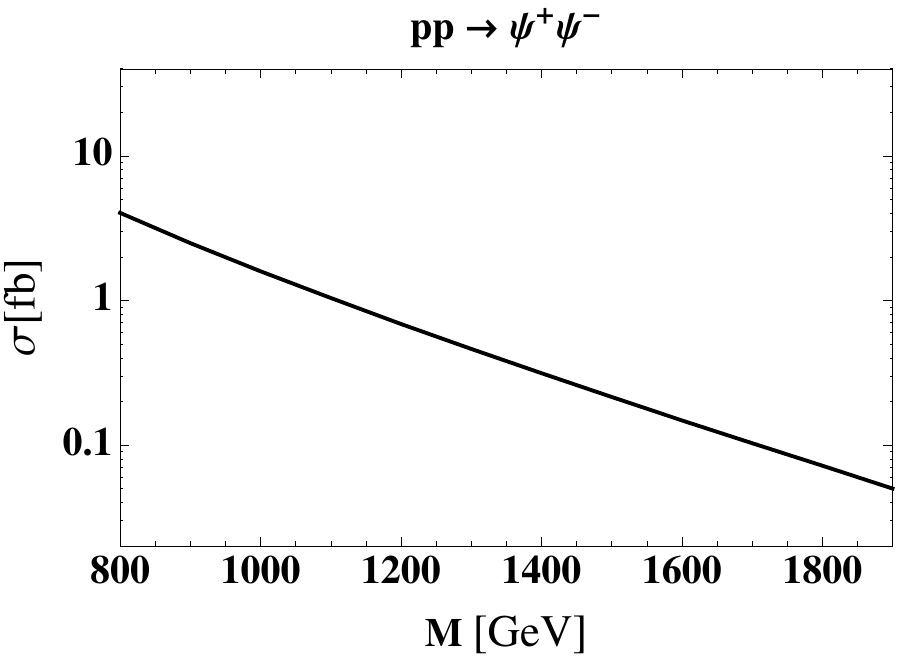}
\caption{Production cross section of $\psi^+ \psi^-$ at the LHC 14 TeV.}   
\label{fig:cx}\end{center}\end{figure}

Fig.~\ref{fig:cx} shows cross section for $pp \to \psi^- \psi^+$ process as a function of $\psi^\pm$ mass where we use {\it CalcHEP}~\cite{Belyaev:2012qa} for estimation with $\sqrt{s} = 14$ TeV.
We obtain $\mathcal{O}(1)$ fb production cross section for TeV scale $\psi^\pm$ at the LHC experiments.

\begin{figure}[t]\begin{center}
\includegraphics[width=80mm]{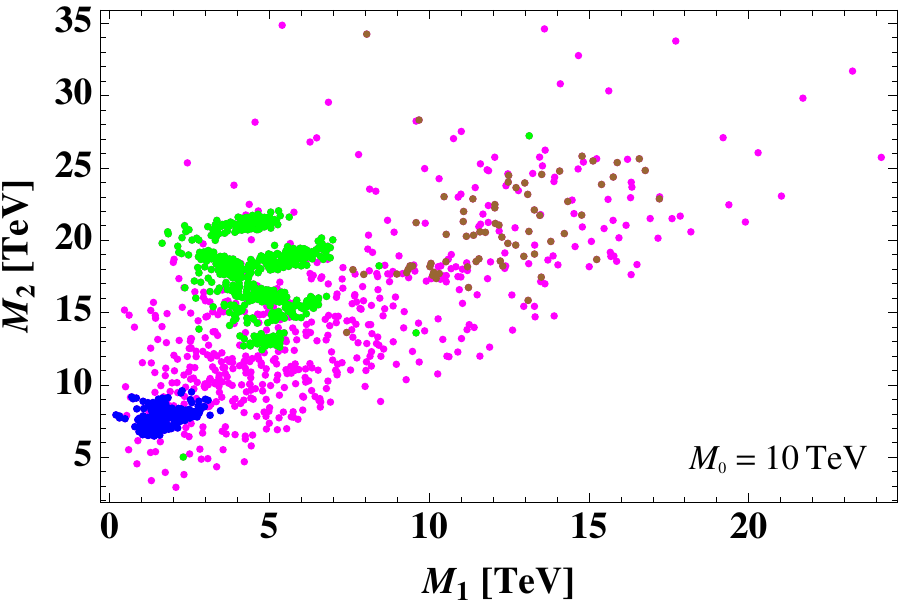}\
\includegraphics[width=80mm]{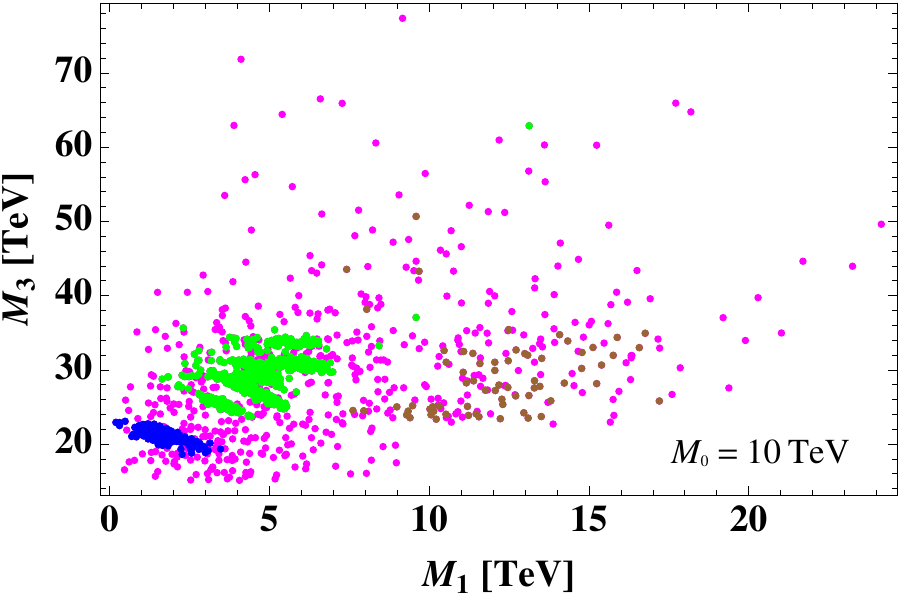}\
\caption{Correlations among masses of three generations of quartet fermion $\psi_4$,
where the color represents the same as previous plots.}   
\label{fig:psi4mass}\end{center}\end{figure}

\begin{figure}[t]\begin{center}
\includegraphics[width=80mm]{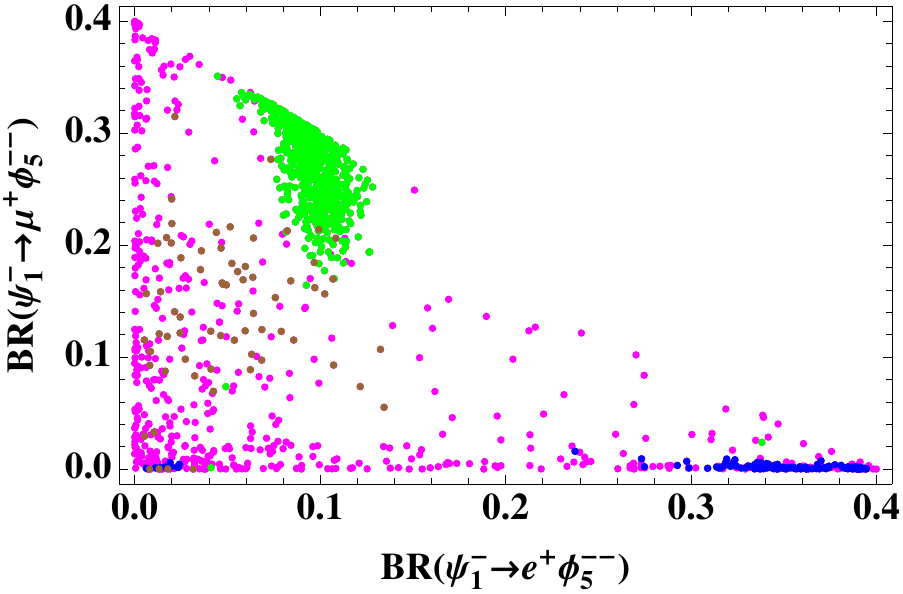}\
\includegraphics[width=80mm]{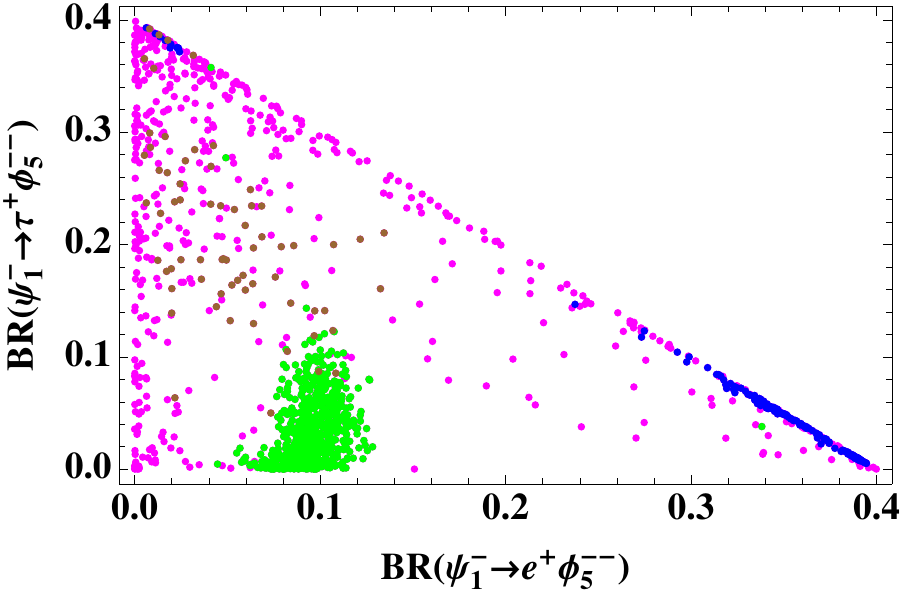}\
\caption{Correlations among branching rations for decay mode $\psi_1^- \to \ell^+ \phi_5^{--}$,
where the color represents the same as previous plots.}   
\label{fig:psi4BR}\end{center}\end{figure}

Here we estimate masses of quartet fermions $\psi_4$ and BR of  $\psi_1^- \to \ell^+ \phi_5^{--}$ modes ($\psi_1^-$ is the first generation of $\psi^-$) applying parameters which can fit neutrino data.
The mass eigenvalues can be obtained by diagonalizing $M$ in Eq.~\eqref{eq:M} ignoring small quantum corrections.
Here BRs of  $\psi_1^- \to \ell^+ \phi_5^{--}$ are given by 
\begin{equation}
BR(\psi_1^- \to \ell^+ \phi_5^{--}) = \frac{| \sum_a (m_D)_{\ell a} V_{a 1}|^2}{\sum_\ell | \sum_a (m_D)_{\ell a} V_{a 1}|^2 + (3/2) \sum_\ell | \sum_a (m_D)_{\ell a} V_{a 1}|^2 }, 
\end{equation}
where $V$ is matrix diagonalizing $M$ and second term in the denominator corresponds to contribution from $\psi_1^- \to \nu \phi_5^{-}$ modes.
In Fig.~\ref{fig:psi4mass}, we show correlations among masses of three generations of quartet fermion $\psi_4$ where different colors correspond to those in previous plots.
It is found that masses tend to be lighter for $\tau=i\times \infty$ fixed point when we determine $M_0$ value.
Also masses are hierarchical for fixed points $\tau=i\times \infty$ and $\tau=i$ while it is less hierarchical for fixed point $\tau = w$. 
In Fig.~\ref{fig:psi4BR} we show correlations among branching rations for decay mode $\psi_1^- \to \ell^+ \phi_5^{--}$. 
We find that decay mode with $e$ or $\tau$ is dominant for fixed point $\tau=i\times \infty$, muonic mode is dominant for fixed point $\tau=i$, 
and all lepton modes have similar BR for  fixed point $\tau = w$.

Here, we pick up benchmark points from the three fixed points $\tau=i\times \infty$,  $\tau=i$ and $\tau = w$ and discuss preferred signals.

\noindent
{\bf Benchmark point I at $\tau=i\times \infty$}:
In this case, $\psi^-_1$ dominantly decays into electron or tau mode and muonic mode is almost absent. 
Thus our signals are  $W^+W^+ W^- W^- e^+ e^-$,  $W^+W^+ W^- W^- \tau^+ \tau^-$ and $W^+W^+ W^- W^- e^\pm \tau^\mp$.

\noindent
{\bf Benchmark point II at $\tau=i$}:
In this case, $\psi^-_1$ dominantly decays into muonic mode. 
Thus our signal is  $W^+W^+ W^- W^- \mu^+ \mu^-$.

\noindent
{\bf Benchmark point III at $\tau=w$}:
In this case, $\psi^-_1$ decays into all lepton mode with similar BRs.
Thus we have signals of $W^+W^+ W^- W^- \ell^+ \ell'^-$ with various lepton final states.
In addition, mass hierarchy of $\psi^-$ generations are small so that we may find second or third generation with the first generation
if $M_0$ value is not very high.

\section{Summary and discussion}

We have constructed an inverse seesaw model with $SU(2)_L$ multiplet fields applying modular $A_4$ symmetry.
The Yukawa couplings for inverse seesaw mechanism are given by modular forms which are functions of modulus $\tau$.
Thus neutrino mass matrix is determined by the modulus and some free parameters. 
 
Then we have numerically analyzed neutrino mass matrix to search for parameters which can fit neutrino data and some predictions.
We have found that some predictions for neutrino sector can be obtained such as CP phases, sum of neutrino mass and effective mass for neutrinoless double beta decay.
In particular, we can predict specific region of them focusing on fixed points for modulus preferred by a string theory although we have several free parameters.
In addition we have discussed collider physics focusing on charged fermion from $SU(2)_L$ quartet since its signal pattern is related to neutrino sector.
We have shown mass hierarchy of three generations of the quartet and branching ratio of $\psi_1^- \to \ell^+ \phi_5^{--}$ decay mode 
applying parameters satisfying neutrino data.
As a result we have obtained some unique patterns of them especially around the fixed points of the modulus.

\section*{Acknowledgments}
\vspace{0.5cm}
{\it
We would like to thank Prof. Sudhanwa Patra for fruitful discussion. 
The work is supported in part by KIAS Individual Grants, Grant No. PG054702 (TN) at Korea Institute for Advanced Study.
This research is also supported by an appointment to the JRG Program at the APCTP through the Science and Technology Promotion Fund and Lottery Fund of the Korean Government. This was also supported by the Korean Local Governments - Gyeongsangbuk-do Province and Pohang City (H.O.). H. O. is sincerely grateful for the KIAS member.}


\end{document}